\documentclass[aps,prb, twocolumn, superscriptaddress, longbibliography, nobalancelastpage]{revtex4-1}

\usepackage{times}
\usepackage{amsmath}
\usepackage{amssymb}
\usepackage{graphicx}
\usepackage[caption=false, position=top, singlelinecheck=off, justification=raggedright]{subfig}
\usepackage{color}
\usepackage{pst-node}
\usepackage[unicode]{hyperref}
\hypersetup{unicode=true, colorlinks=true, linkcolor=black, citecolor=black,urlcolor=black }
\usepackage{float}
\usepackage[normalem]{ulem}
\usepackage{multirow}
\usepackage{hhline}

\usepackage{mathrsfs,braket}
\usepackage{amssymb, amsbsy, amsmath, latexsym, dsfont, array, layout, graphicx, mathrsfs, color, ulem, bm}

\usepackage{changepage}  

\usepackage{xcolor}
\usepackage{tikz}
\definecolor{ACSblue}{RGB}{38,51,128}

\usepackage[T1]{fontenc}
\usepackage[scaled=1.0]{sourcesanspro}

\newcommand{\ACSSISubtitle}[1]{
  {
\fontfamily{SourceSansPro-TLF}\fontseries{sb}\selectfont
   \normalsize #1}
}
 

\makeatletter
\renewcommand*{\fnum@figure}{\textbf{Figure \thefigure}}

\renewcommand*{\@caption@fignum@sep}{\textbf{.} }
\makeatother

\definecolor{orcidlogocol}{HTML}{A6CE39}

\begin{document}


\title{What Is the Real-Time Atomistic Mechanism Behind Chirality-Induced Spin Selectivity in Donor-Chiral Bridge-Acceptor Molecules? }

\author{Shu-Zheng Zhou}
 \affiliation{School of Physics and Wuhan National High Magnetic Field Center,
Huazhong University of Science and Technology, Wuhan 430074, People's Republic of China.}

\author{Xi Sun}
 \affiliation{School of Physics and Wuhan National High Magnetic Field Center,
Huazhong University of Science and Technology, Wuhan 430074, People's Republic of China.}

\author{Kai-Yuan Zhang}
 \affiliation{School of Physics and Wuhan National High Magnetic Field Center,
Huazhong University of Science and Technology, Wuhan 430074, People's Republic of China.}

\author{Hua-Hua Fu}
\altaffiliation{Corresponding author.\\ hhfu@hust.edu.cn}
\affiliation{School of Physics and Wuhan National High Magnetic Field Center,
Huazhong University of Science and Technology, Wuhan 430074, People's Republic of China.}
\affiliation{Institute for Quantum Science and Engineering, Huazhong University of Science and Technology, Wuhan, Hubei 430074, China.}

\date{\today}

\begin{abstract}
\noindent
Chiral-induced spin selectivity (CISS) has been experimentally observed in photo-excited donor–chiral bridge–acceptor (D‑B$\chi$‑A) molecules [Science 382, 197–201 (2023)]. However, the microscopic mechanism underlying CISS in such chiral systems remains elusive. Here we develop a quantum dynamical model that precisely maps the atomic structure of binaphthyl-type bridge dimers in isolated D‑B$\chi$‑A molecules and introduce a geometric spin–orbit coupling (SOC) mechanism to unveil the intrinsic origin of CISS in axially chiral systems. During photo-excited electron transport along the twisted pathways, the geometric SOC coupling strength exceeds the intrinsic coupling of light atoms by one to two orders of magnitude, readily producing observable high spin polarizations. The resulting spin polarization comprises two components: the CISS-associated polarizations along and perpendicular to the chiral axis are intrinsic to axial chirality, requiring neither external fields nor spin-superexchange transfer, while a non-Abelian curvature correction provides a rigorous mathematical definition of the chiral axis direction. Our calculated polarization components, chirality dependence, and relative magnitudes (30\%–40\%) quantitatively match time-resolved electron paramagnetic resonance measurements. This geometric SOC framework offers a self-consistent and general physical picture of CISS in axially chiral molecules and provides explicit theoretical guidance for the design of chiral spintronic devices.

\end{abstract}
\maketitle


\noindent The generation of spin-polarized currents by nonmagnetic chiral molecules in the absence of external magnetic fields, i.e., the chirality-induced spin selectivity (CISS) effect \cite{1,2,3,4}, still has laced a unifying mechanistic framework since its discovery in DNA \cite{7,8,9,10,11,12,13,14,15}. Widespread experimental verification across various chiral systems \cite{7,8,9,10,11,12,13,14,15} has done little to settle the debate \cite{16,17,18}, as the vast majority of transport measurements rely on metallic or ferromagnetic contacts whose heavy-atom spin-orbit coupling (SOC) obscures intrinsic chiral molecular contributions \cite{19,20,21,22}. This persistent confounding factor has now been circumvented by Eckvahl \textit{et al.} through time-resolved electron paramagnetic resonance measurements on isolated donor–chiral bridge–acceptor (D‑B$\chi$‑A) molecules \cite{23,24,25}. Devoid of substrate- and electrode-induced artifacts, their data directly trace ultrafast charge and spin separation as electrons traverse the chiral bridge, thereby rigorously refocusing the quest for mechanism onto the chiral molecular itself.

Yet the same measurements also expose several unconventional spin transport features \cite{23,24,25}. The spin population accumulated on the acceptor moiety consists of both chirality-associated and chirality-independent contributions. More strikingly, the CISS signals persists irrespective of molecular orientation, whether in isotropic solution or aligned within liquid crystals, indicating that polarization vector is not confined to the chiral axis but possesses a substantial orthogonal component. No existing theoretical treatment simultaneously account for both sources of polarization \cite{26,27,28,29}. The one-dimensional helical chain model, applicable only to helically symmetric systems, projects spin exclusively along the helix axis and therefore inherently precludes any transverse contribution \cite{28}. The phenomenological two-level model \cite{26,27}, although heuristically valuable for interpreting time-resolved spectra, omits the real-time dynamics of polarization buildup and neglects the detailed architecture of the chiral bridge unit. A recent extension that incorporates low-frequency torsional vibrations to modulate electron hopping and SOC introduces an effective Dzyaloshinskii–Moriya (DM) interaction to rationalize the perpendicular components \cite{29}; its predictions, however, depend sensitively on narrowly constrained vibrational frequencies, coupling constants, and charge-transfer rates, which severely curtail its general validity. Critically, all these formulations bypass the explicit role of transport pathways and the fine structural parameters of chiral bridge, specifically, the dihedral angles between naphthyl rings, the interplanar spacings, and the donor/acceptor anchoring sites, that may govern the observed polarization characteristics.

What remains unexplained, therefore, is not merely the coexistence of two polarization components in an axially chiral system, but the mechanistic origin of the transverse CISS contribution itself and its parametric dependence on the molecular scaffold. Does the perpendicular polarization arise from a purely electronic effect, such as SOC mediated by non-adiabatic coupling between electronic states, or does it require the participation of low-frequency nuclear motions that break the axial symmetry? More broadly, which specific geometric and electronic descriptors, the dihedral twist angle, or the inter-ring coupling strength, most sensitively control the net spin polarization yield? The 2023 Science report provided unambiguous evidence for CISS in these D‑B$\chi$‑A systems and pointed to their prospective utility in quantum information processing \cite{23}, yet it left entirely open the central question of how to rationally engineer the chiral bridge to maximize or switch the spin polarization at the molecular level. Addressing this requires not only systematic variation of bridge geometry and chemical composition, but also a theoretical framework that goes beyond one-dimensional transport and vibronic corrections to treat the full three-dimensional nature of molecular potential and its coupling to the spin degree of freedom.

\begin{figure*}[t]
\includegraphics[width=2.0\columnwidth]{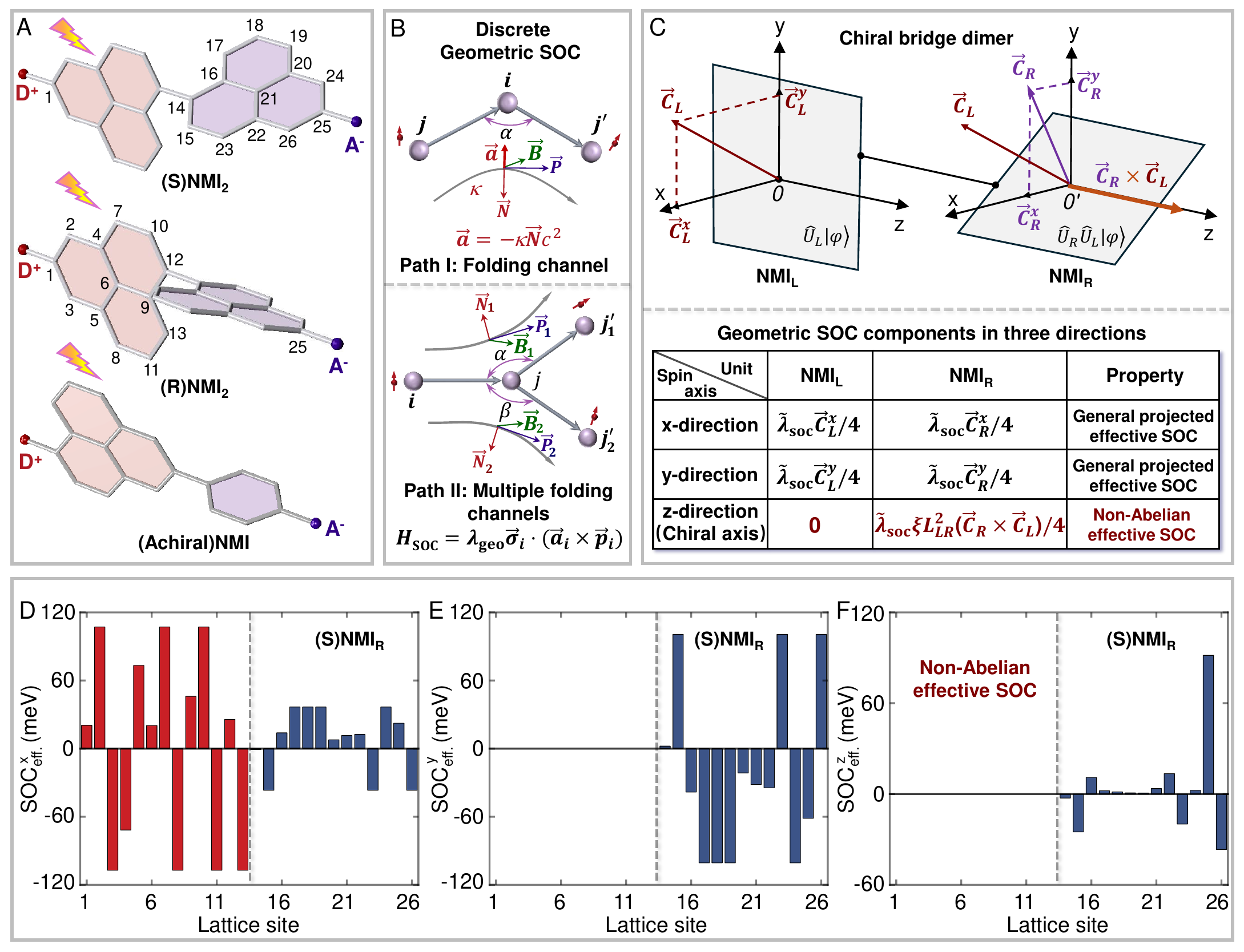}
\caption{\textcolor{black}{\textbf{Molecular geometry of the donor–chiral bridge–acceptor system and the generation of geometric SOC within the chiral bridge.} \textbf{A} Molecular structures of the PXX-(S,R)NMI$_2$–NDI enantiomers, the achiral NMI-bridged analogue is also shown for reference. \textbf{B} Two representative transport pathways through the chiral bridge and the discretized geometric SOC at each atomic site. The vectors $\vec{N}$, $\vec{B}$, and $\vec{T}$ denote the normal, binormal, and tangent directions in the Frenet–Serret frame; $\vec{a}$ and $\vec{p}$ are the electron acceleration and momentum; $\kappa$ is the local curvature of the pathway. \textbf{C} Decomposition of the effective geometric SOC components along the three orthogonal axes for the left (NMI$_{\rm L}$) and right (NMI$_{\rm R}$) units of the chiral-bridge dimer, $C_L$ and $C_R$ are the quantum-current-weighted accumulated curvature vectors for the two units, respectively. \textbf{D-F} Spatial distribution of the effective geometric SOC components $\rm SOC^{x,y,z}_{eff}$ along the atomic sites of NMI$_{\rm L}$ ($i$=1-13, red) and NMI$_{\rm R}$ ($i$=1-13, orange) for the (S) enantiomer.}}
\label{fig1}
\end{figure*}

To address these questions, we investigate the PXX‑(S,R)NMI$_2$‑DNI molecule, focusing on its axially chiral binaphthyl-type bridge dimer with full atomistic resolution. We construct a three-dimensional, real-space time-dependent quantum dynamics model that explicitly incorporates actual nuclear coordinates of the chiral bridge and the donor-acceptor connection geometry. At the core of our approach is a geometric SOC that arises from the curvature of the electron transport trajectory along the twisted intramolecular pathway \cite{30,31,32}; its magnitude at each atomic site is determined by the local torsion angle and interplanar spacing, thereby establishing a direct, parameter-free mapping between molecular conformation and SOC strength. This geometric contribution, which scales with the gradient of the electron momentum along the curved path, typically exceeds the intrinsic SOC of light atoms by one to two orders of magnitude \cite{33,34}. We evaluate the SOC matrix elements from the real-space atomic positions and local symmetry operations, and then propagate the time-dependent charge-state populations, spin densities, and their spatial orientation across the bridge dimer. Our calculations reproduce both the longitudinal and transverse components of spin polarization in the absence of external fields or auxiliary spin-manipulation schemes. Notably, the transverse polarization resolves into a CISS-correlated fraction ($\sim$30–40\% of the total) and a residual CISS-uncorrelated background, a decomposition that we achieve by comparing the polarization dynamics under opposite enantiomers and by isolating the symmetric contribution under time reversal. The geometric SOC, governed by the twisted transport pathway, combined with the asymmetry of the donor-to-acceptor hoppings, uniquely accounts for both polarization sources within a single, coherent framework, thus providing a unified physical picture that reconciles the apparent dichotomy between axial chirality and transverse spin selectivity.\\

\noindent{\bfseries Generation of geometric SOC in D‑B$\chi$‑A molecules}\\
\noindent Our calculations are performed on the D‑B$\chi$‑A enantiomers, (R)‑1‑$\rm h_9$(‑$\rm d_9$) and (S)‑1‑$\rm h_9$(‑$\rm d_9$), employed in the experiments of Eckvahl et al \cite{23}. The donor is peri‑xanthenoxanthene (PXX, protiated or fully deuterated), the acceptor is naphthalene‑1,8:4,5‑bis(dicarboximide) (NDI). The chiral bridge B$\chi$ is a binaphthyl-type NMI$_2$ dimer whose (S)/(R) handedness defines the axial chirality. The inter-ring dihedral angle and the interplanar separation of two NMI units determine the curvature of the electron transport trajectory, hence the geometric SOC is produced. The full non-hydrogen skeleton of the bridge contains 28 atoms, and we retain all atom coordinates from the optimized ground-state geometry (Figure 1A). As a control, we replace the chiral dimer with an achiral DMI bridge (19 atoms) to isolate the axial chirality effect. The donor and acceptor are treated as two-level quantum dots, each with a highest occupied and lowest unoccupied molecular orbital; photoexcitation elevates an electron from the donor ground state to its excited state, after which the electron traverses the chiral bridge and localizes on the acceptor.

In the absence of helical symmetry or periodic spin–orbit potentials in D‑B$\chi$‑A molecules, we introduce a geometric spin–orbit coupling (SOC) mechanism: as the photo-excited electron moves along the twisted path within the chiral bridge, its momentum direction continuously changes with the local curvature, producing an effective magnetic field on the electron spin and thereby inducing spin polarization. The continuous form of this mechanism can be described in the Frenet–Serret frame. Let the tangent and normal vectors at the $i$th site be $\vec{T}_i$ and $\vec{N}_i$, respectively; the local effective acceleration of the electron is $\vec{a}_{i}$=$-\kappa_{i} c^2 \vec{N}_{i}$, where $\kappa_{i}$ is the local curvature and $c$ is the speed of light. The corresponding geometric SOC Hamiltonian reads $\mathcal{H}_{\mathrm{geo}}$=$\hbar (\vec{a}\times\vec{p})\cdot \vec{\sigma}/{(4mc^{2})}$, where $\vec{p}$ is the momentum, $m$ is the electron mass, and $\vec{\sigma}$ is the Pauli vector. The curvature $\kappa_{i}$ is computed directly from the atomic coordinates of the bridge dimer, linking the SOC strength explicitly to molecular conformation and closely related to the molecular conformation. Noted that a related effect, curvature‑induced SOC, has been experimentally observed in carbon nanotubes, serving as a manifestation of geometric SOC in curved quasi‑one‑dimensional systems \cite{35}.

Discretizing the geometric SOC onto the atomic sites of the non-periodic bridge yields, in the Frenet–Serret frame,
\begin{equation}
\mathcal{H}_\mathrm{geo}= - \sum_i \frac{\hbar \kappa_i}{4 m} (\vec{N}_i \times \vec{p}_i)\cdot \vec{\sigma},
\end{equation}
Symmetrization to enforce Hermiticity gives $\mathcal{H}_{\rm geo} = \sum_i \frac{\hbar}{8 m} \kappa_i (p_\parallel \cdot \sigma_i + \sigma_i \cdot p_\parallel)$, with $\sigma_i = \vec{\sigma}\cdot\vec{B}_i$. After second quantization and with the hopping integral $t_0 = 2.7\ {\rm eV}$, the lattice Hamiltonian for the real molecule becomes
\begin{equation}
\mathcal{H}_\mathrm{G-SOC}=i \frac{\lambda}{2} \sum_i \sum_{j\in\langle i\rangle}  \frac{c_j^\dagger}{n_i-1} \sum_{j'\in\langle i\rangle, j'\ne j} \frac{\kappa_{j i j'}\, \sigma_{j i j', B}}{l_{j i j'}} c_i + {\rm H.c.},
\end{equation}
where $\lambda$=$\hbar^2/(4m) \approx 1.7\ {\rm eV}\cdot$\AA$^2$, $n_i$ is the coordination number of the $i$th atom, and $l_{j i j'}$ is the sum of bond lengths along the path $j$$\to$ $i$$\to$$j'$. The The curvature for a single path is $\kappa_{j i j'}$=$4\cos(\alpha/2)/l_{j i j'}$, with $\alpha$ is the bond angle and $l_{j i j'}$=$l_{ji} + l_{i j'}$; for branching paths, the effective curvature is weighted by the electron flux (Figure 1B). The ratio $\lambda/t_0\approx 0.64$  (see Supplementary Information for derivation) sets the intrinsic SOC energy scale in our tight-binding framework (see Methods).\\

\noindent{\bfseries The key contribution of non-Abelian SOC to CISS}\\
\noindent To elucidate the physical origin of spin polarization following photoexcited electron transfer through the chiral bridge, we first examine the influence of the geometric configuration of the NMI$_2$ dimer on the transferred electron. In the model, all atoms in the left and right units NMI$_{\rm L}$ and NMI$_{\rm R}$ are kept coplanar, with bond lengths and angles consistent with the actual molecule; the dihedral angle between the two units is set to 70$^\circ$ and -70$^\circ$, matching the experimental molecule (Figure 1A). A Cartesian coordinate system is established with reference to NMI$_{\rm L}$: the $x$ direction is perpendicular to the NMI$_{\rm L}$ plane, the $z$ direction is along the chiral axis, and NMI$_{\rm L}$ lies in the $yz$ plane; the spatial positions of NMI$_{\rm R}$ and acceptor A are referenced to this frame.

\begin{figure*}[t]
\includegraphics[width=2.00\columnwidth]{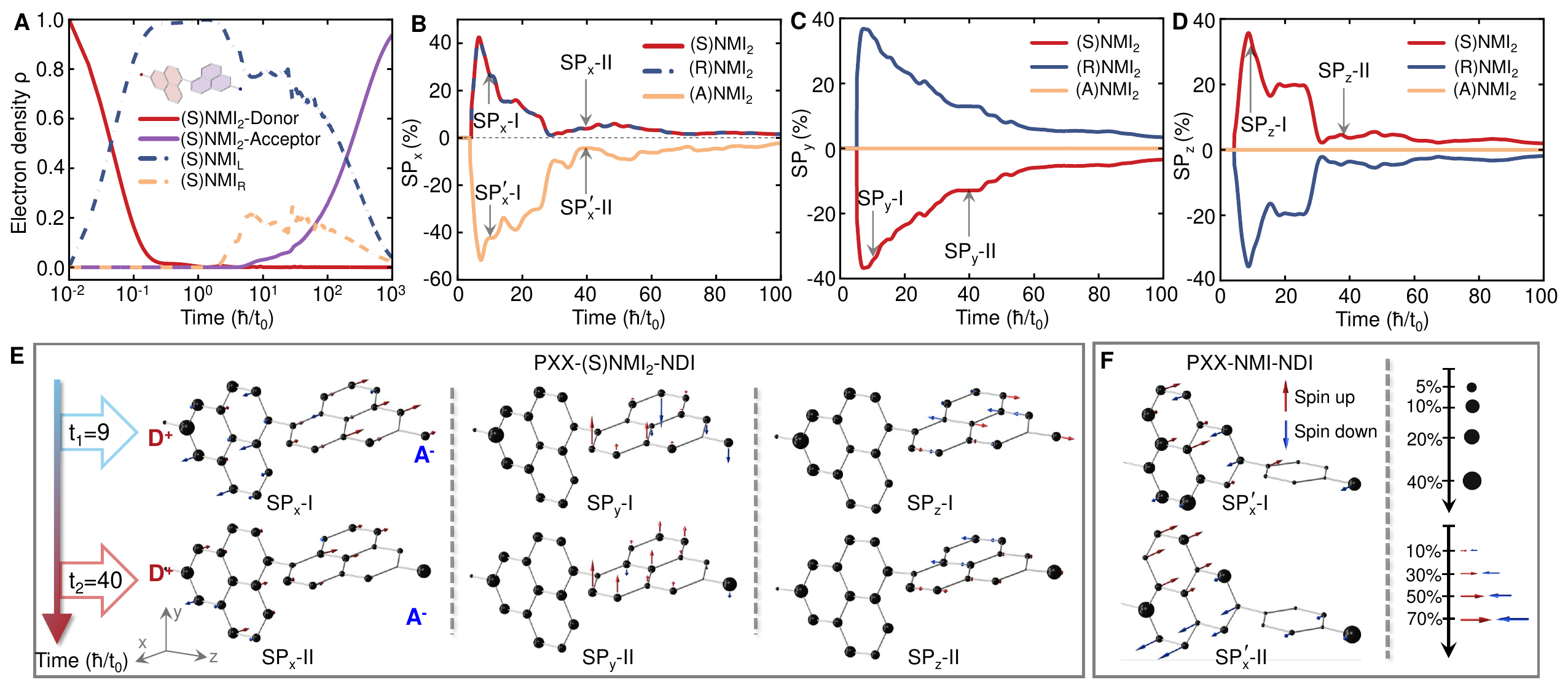}
\caption{\textbf{Time and spatial evolution of charge densities and spin polarizations in the donor–chiral bridge–acceptor system.} (\textbf{A}) Time-dependent charge densities on the donor (D), the left bridge unit NMI$_{\rm L}$ (adjacent to D), the right bridge unit NMI$_{\rm R}$ (adjacent to the acceptor A), and the acceptor itself. (\textbf{B-D}) Time evolution of the three orthogonal spin-polarization components $SP_x$, $SP_y$, and $SP_z$ on the acceptor for the (S) and (R) enantiomers and the achiral reference. (\textbf{E}) Spatial distribution of charge densities and the three spin-polarization components across all atomic sites of the chiral-bridge (S)NMI$_2$ and the acceptor at two representative times, $t_1$ and $t_2$. (\textbf{F}) Corresponding spatial distribution for the achiral PXX–NMI–NDI molecule, showing charge densities and the $SP_x$ component at the above times.}
\label{fig2}
\end{figure*}

After transfer to the chiral bridge, the electron moves along a twisted pathway and experiences geometric SOC at each atomic site, generating spin polarization. We computed the effective SOC components (denoted as $\rm SOC^x_{\rm {eff}}$, $\rm SOC^y_{\rm {eff}}$, $\rm SOC^z_{\rm {eff}}$) in three orthogonal directions for each site of NMI$_{\rm L}$ and NMI$_{\rm R}$ (Figure 1D-1F). The results reveal that (i) $\rm SOC^x_{\rm {eff}}$ is finite at all sites of both units, with significantly larger magnitude on NMI$_{\rm L}$ and that on NMI$_{\rm R}$; (i) $\rm SOC^y_{\rm {eff}}$ and $\rm SOC^z_{\rm {eff}}$ are strictly zero on  NMI$_{\rm L}$ but appear on NMI$_{\rm R}$. This indicates that spin polarization along $x$ direction exists in both units, whereas polarizations along $y$ and $z$ directions originate exclusively from the right unit, and the distribution is intimately related to the molecular conformation and electron transport pathway, reflecting the dynamic transport character of geometric SOC.

The emergence of effective SOC components along $y$ and $z$ directions in the right unit NMI$_{\rm R}$ is attributed to the difference in non-Abelian effective curvature contributions between the left and right units. The photoexcited electron first accumulates SOC in the left half‑plane, and upon entering the right half‑plane, it further accumulates a second‑order curvature correction arising from non‑collinear SOC parameters along the actual path. To quantify the influence of this non‑Abelian term on spin polarization, we identify the transport pathways sampled by the wave packet via the Hamiltonian bond probability current $J_{i\to j}^H(t) = \frac{2}{\hbar} \mathrm{Im}\,\mathrm{Tr}_s [\mathcal{H}_{ij}\rho_{ji}(t)]$. The terminal source/drain currents are evaluated by Lindblad hopping fluxes, and the time‑integrated effective net flow defines the direction and weight of each transport edge. A quantum‑current‑weighted effective curvature is assigned to each site by averaging the local spin‑projected curvature over all incoming–outgoing corners. On this basis, we introduce a non‑Abelian correction for the left/right planes, which originates from the path‑ordered spin evolution under non‑collinear SOC fields. For two successive SOC rotations ($U_L$ and $U_R$), the Baker–Campbell–Hausdorff expansion yields $U_R U_L \approx \exp[-i (a_R + a_L + a_R \times a_L)\cdot \sigma]$, indicating that non‑collinear SOC rotations generate a second‑order non‑Abelian curvature proportional to the cross product of the accumulated curvature vectors. This term is explicitly evaluated as $K_{LR}^{(2,\mathrm{NA})}$ = $L_{LR}\, \hat{n} \cdot ( \mathbf{C}_R \times \mathbf{C}_L )$, or in the expand form
\begin{equation}
K_{LR}^{(2,\mathrm{NA})} = L_{LR} \sum_{a\in L}\sum_{b\in R} \omega_a^L \omega_b^R\, \hat{n} \cdot [ (\kappa_b \hat{B}_b) \times (\kappa_a \hat{B}_a) ].
\end{equation}
where $\mathbf{C}_L$ and $\mathbf{C}_R$ are the quantum‑current‑weighted accumulated curvature vectors on the left and right molecular planes, respectively. This term captures the non-commuting spin-rotation history experienced by the transporting electron, which is absent in the Abelian projected SOC approximation; the cross product corresponds to the time order ``left then right'', and reversing the transport direction or swapping the order of operations changes its sign. This non-Abelian correction generates an additional projected component of SOC along the chiral axis within the NMI$_{\rm R}$ plane. This nonzero axial geometric SOC component arises solely from the torsional dihedral angle of NMI$_{\rm R}$ relative to NMI$_{\rm L}$ and is a direct manifestation of the inherent spatial structure of axial chirality. For the enantiomer (R)NMI$_2$, the $x$‑component of SOC remains the same as in the (S) form, while the $y$ and $z$ components have equal magnitude but opposite sign (particularly the non‑Abelian term), further confirming the axial chirality character and demonstrating that one‑dimensional helical‑chain models are inapplicable to CISS studies in D‑B$\chi$‑A type molecules.

In summary, the special geometric configuration of the chiral bridge gives rise to two sources of spin polarization from the geometric SOC experienced by the transferred electron: one originates from the twisted pathways within each of the two NMI$_2$ units and is independent of axial chirality; the other arises from the twisted pathways between the two units and is intimately associated with axial chirality, constituting the key mechanism for the CISS effect. According to our model, when the two bridge units become coplanar (i.e., axial chirality vanishes), the $x$‑direction spin polarization persists, while the $y$‑ and $z$‑direction polarizations disappear. This confirms that two distinct spin‑polarization components indeed coexist in the PXX-NMI$_2$–NDI molecule.\\ 

\noindent{\bfseries Evolution of charge and spin-polarization dynamics}\\
\noindent We first examine the real-time charge transfer in photoexcited PXX–(S)NMI$_2$–NDI molecule. Figure 2A shows the time evolution of charge densities $\rho$ on the donor (D), acceptor (A), left bridge unit NMI$_{\rm L}$ (adjacent to D), and right bridge unit NMI$_{\rm R}$ (adjacent to A). Immediately after citation ($t$<$10^{-2}\hbar/t_0$), the charge resides almost entirely on D. The density at the leftmost site of NMI$_{\rm L}$ then rises sharply, peaking at $t\approx{\hbar/t_0}$, while the donor density decays correspondingly. At $t\approx5{\hbar/t_0}$, the leftmost site begins to deplete, and the rightmost site of NMI$_{\rm R}$ accumulates charge, but with a peak roughly one order of magnitude lower and delayed relative to the left side. This marked asymmetry, fast injection into the left unit versus slow, attenuated arrived at the right unit, indicates that the electron experiences substantial geometric SOC at each atomic site along the twisted bridge, which impedes coherent transmission and promotes charge–spin separation. the charge densities at the terminal bridge atoms (B$_1$ and B$_{\rm N}$) oscillate with multiple transient peaks, confirming that geometric SOC acts locally throughout the pathway. By $t\approx10{\hbar/t_0}$, acceptor density begins to rise, reaching a level comparable to the initial donor population at $t\approx10^3{\hbar/t_0}$, marking the completion of the majority of charge transfer.

We next examine the temporal behavior of the three orthogonal spin-polarization components on the acceptor, $\rm SP_x$, $\rm SP_y$, and $\rm SP_z$, for the (S) enantiomer, its (R) counterpart, and an achiral NMI-bridged reference (Figures 2B-2D). For the (S) enantiomer, all three components peak at approximately 40\% near $t\approx10{\hbar/t_0}$, then decay gradually. The polarization maximum precedes the charge-density peak on the acceptor by nearly two orders of magnitude, a feature qualitatively captured by time-resolved electron paramagnetic resonance (TREPR) experiments. Across the two enantiomers, $\rm SP_y$ and $\rm SP_z$ are nearly equal in magnitude but opposite in sign, whereas $\rm SP_x$ retains the same sign; this sign pattern directly mirrors the directional distribution of the geometric SOC components, identifying $\rm SP_y$ and $\rm SP_z$ are chirality-dependent and as chirality-independent. The calculated chirality-related fraction (30-40\% of the total polarization) agrees quantitatively with the two-component decomposition reported in the TREPR spectra of Eckvahl et al. The geometric SOC framework thus accounts naturally for the coexistence of two distinct polarization classes within a single chiral molecule, a distinction that previous two-level or helical-chain models could not produce. Corroborating this assignment, the achiral bridge system yields finite $\rm SP_x$ but strictly zero $\rm SP_y$ and $\rm SP_z$, confirming that only the transverse components are selective to molecular handedness.

We then turn to the real-space distribution of spin polarization at two representative times, $t_1=9{\hbar/t_0}$ and $t_2=40{\hbar/t_0}$, analyzing the instantaneous charge densities and polarization vectors on the donor, the NMI$_2$ bridge, and the acceptor (Figure 2E, six spin-polarized states $\rm SP_{x,y,z}$-${\rm I, \rm {II}}$). At $t_1$, charge resides predominantly on NMI$_{\rm L}$, with a smaller population on NMI$_{\rm R}$ that decays along the transfer direction. By $t_2$, NMI$_{\rm R}$ and the acceptor acquire substantial densities, with higher populations at centrosymmetric bridge sites than at terminal ones, reflecting an asymmetric transport pathway. Except for the donor, all bridge sites and the acceptor exhibit significant $x$-direction polarization. Notably, polarization emerges on bridge sites as soon as they become populated, well before the donor is fully depleted, accounting for the early polarization peak relative to charge transfer. Sites along single-pathway channels (e.g., molecular edges) display markedly higher polarization than those at bifurcated junctions (e.g., near-center atoms with multiple bonding neighbors), indicating that pathway topology modulates the effective geometric SOC strength. Together, these observations establish that efficient spin polarization in this system requires not only the geometric SOC arising from the twisted bridge, but also an asymmetric charge-transfer pathway that enhances the local curvature effect along the dominant transport route.

Unlike $\rm SP_x$, which appears on both NMI units, $\rm SP_y$ and $\rm SP_z$ are strictly confined to the NMI$_{\rm R}$ moiety and the acceptor, vanishing entirely on NMI$_{\rm L}$. Thus, the right half of the chiral bridge serves as the exclusive spatial origin of the chirality-dependent polarization components, while the left half contributes only to the chirality-independent $\rm SP_x$. This real-space segregation directly corroborates the component classification inferred from the enantiomer sign patterns. Calculations for the (R) enantiomer confirm that $\rm SP_y$ and $\rm SP_z$ reverse sign while $\rm SP_x$ does not (Figures 2C-2D), and the achiral bridge system yields finite $\rm SP_x$ on both NMI units but strictly zero transverse components (Figure 2F). Together, these findings establish that the emergence of $\rm SP_y$ and $\rm SP_z$ is dictated entirely by the axial chirality of the bridge, with the geometric conformation of the NMI$_{\rm R}$ unit, specifically its connection to the acceptor, playing the primary structural role.\\

\begin{figure*}[t]
\includegraphics[width=1.5\columnwidth]{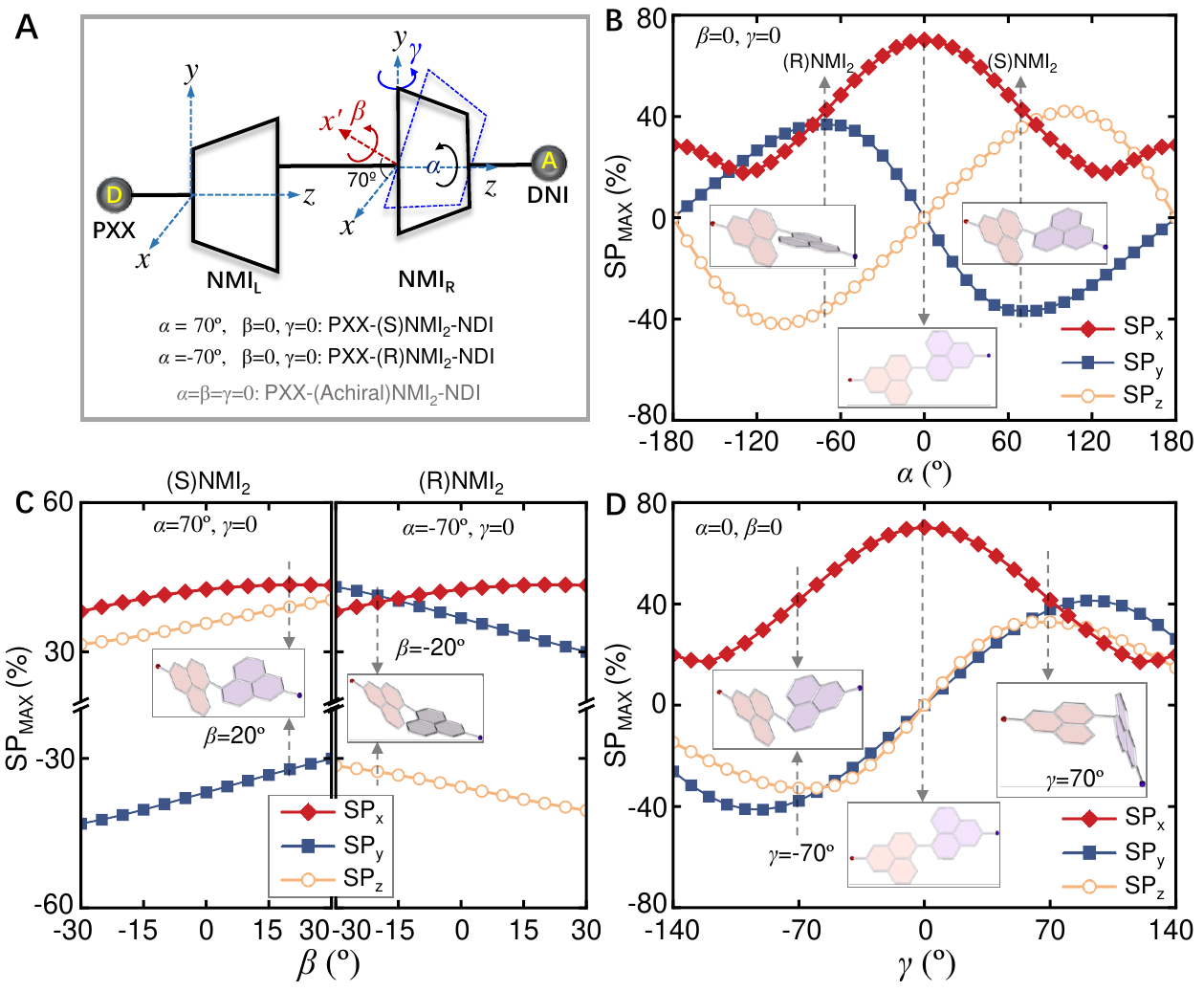}
\caption{\textbf{Geometric regulation of spin polarization by chiral bridge dimer conformation.} (\textbf{A}) Definition of the three geometric parameters of the right NMI$_{\rm R}$ unit relative to the left NMI$_{\rm L}$ unit: dihedral angle $\alpha$ about the chiral axis, in-plane twist $\beta$, and folding angle $\gamma$; the corresponding rotational axes are indicated. (\textbf{B}) Dependence of the three spin-polarization components $\rm SP_x$, $\rm SP_y$, and $\rm SP_z$ on the dihedral angle $\alpha$ for the PXX-NMI$_2$-DNI system. (\textbf{C}) Evolution of the three polarization components with the in-plane twist angle $\beta$ for the enantiomeric pair at $\alpha$=$\pm70^{\circ}$. (\textbf{D}) Variation of the three spin-polarization components on the acceptor A as a function of the folding angle $\gamma$.}
\label{fig3}
\end{figure*}

\noindent{\bfseries Regulation of the CISS by chiral-bridge geometric configuration.}\\
\noindent We next probe the geometric determinants of spin polarization by systematically varying the relative conformation of the two NMI units in the chiral bridge. With NMI$_{\rm L}$ fixed in the $yz$ plane, three parameters of NMI$_{\rm R}$ are independently tuned: the ddihedral angle $\alpha$ about the chiral axis, the in-plane twist $\beta$, and the folding angle $\gamma$ (Figure 3A). The enantiomeric (S) and (R) configurations correspond to $\alpha$=$\pm70^\circ$ with $\beta$=$\gamma$=0; the achiral coplanar arrangement is recovered at $\alpha$=0.

\noindent \textbf{Dihedral angle $\alpha$ variation.} As $\alpha$ sweeps from -$180^\circ$ to $180^\circ$ with $\beta$=$\gamma$=0, $\rm SP_x$ remains positive and symmetric about $\alpha$=0, peaking at the coplanar configuration ($\alpha$=0) and taking identical values at $\alpha$=$\pm70^\circ$ (Figure 3B). In contrast, $\rm SP_y$ and $\rm SP_z$ are antisymmetric about $\alpha$=0, vanish at the achiral limit, and reverse sign upon enantiomeric interchange. This sign pattern identifies $\rm SP_y$ and $\rm SP_z$ as the intrinsic CISS contributions: $\rm SP_z$ along the chiral axis and flips with handedness, while $\rm SP_y$, though transverse, shares the same chirality-dependent behavior, a feature inherent to axial chirality, not requiring extrinsic mechanisms such as Dzyaloshinskii–Moriya interactions. The coexistence of chirality-independent ($\rm SP_x$) and chirality-dependent ($\rm SP_y$ and $\rm SP_z$) transverse components thus accounts for the two polarization sources resolved in the TREPR spectra. This distinction, which emerges naturally from the three-dimensional bridge conformation and site-specific transport pathways, is not captured by one-dimensional helical-chain or simplified two-level models.

\noindent \textbf{In‑plane twist angle $\beta$ variation.} With $\alpha$=$\pm70^\circ$ and $\gamma$=0 fixed, $\beta$ is swept from –30$^{\circ}$ to 30$^{\circ}$. For both enantiomers, the magnitudes of $\rm SP_x$, $\rm SP_y$, and $\rm SP_z$ vary, but their polarization directions remain unchanged (Figure 3C), indicating that $\beta$ tuning does not alter the chirality type. $\rm SP_x$ remains equal for the two enantiomers; $\rm SP_y$ and $\rm SP_z$ retain opposite signs, confirming the classification established from the $\alpha$ scans. As $\beta$ increases from negative to positive, $\rm SP_x$ and $\rm SP_z$ both grow in magnitude, while $\rm SP_y$ weakens. This divergence reflects the changing projection of the effective SOC vector onto the three axes as the in-plane twist modifies the local curvature distribution along the NMI$_{\rm R}$ transport pathway: enhanced torsional curvature amplifies the geometric SOC, but its vector orientation shifts such that the transverse chiral component ($\rm SP_y$) is suppressed while the axial ($\rm SP_z$) and chirality-independent ($\rm SP_x$) components are reinforced. The sensitivity of $\rm SP_y$ to $\beta$ suggests that the detection of transverse CISS signals in TREPR experiments requires specific conformational conditions, a prediction that can be tested by systematic variation of the bridge geometry.

\noindent \textbf{Folding angle $\gamma$ variation.} Starting from the achiral planar reference ($\alpha$=$\beta$=0), we vary $\gamma$ from -135$^{\circ}$ to 135$^{\circ}$. $\rm SP_x$ remains symmetric about $\gamma$=0 and positive, consistent with the planar geometry at $\gamma$=0. In contrast, $\rm SP_y$ and $\rm SP_z$ vanish at $\gamma$=0 and reverse sign simultaneously upon sign change of $\gamma$, exhibiting antisymmetric distributions (Figure 3D). Thus, a nonzero $\gamma$ generates axial chirality in the otherwise achiral framework: the conformations at $\gamma$=+70$^{\circ}$ and $\gamma$=-70$^{\circ}$ form an enantiomeric pair with the chiral axis oriented along $y$ axis. In this pair, $\rm SP_x$ remains chirality-independent, while $\rm SP_y$ and $\rm SP_z$ are strictly tied to the newly created handedness. The vector direction of the non-Abelian geometric SOC term, which originates from the curvature-induced effective field, defines the chiral axis: for $\alpha$ variation, this term aligns with $z$ axis; for $\gamma$ variation, it aligns with $y$ axis. For a generic axially chiral molecule, therefore, the chiral axis is uniquely determined by the orientation of this non-Abelian vector, rather than by any predefined molecular coordinate.\\

\begin{figure*}[t]
\includegraphics[width=2.0\columnwidth]{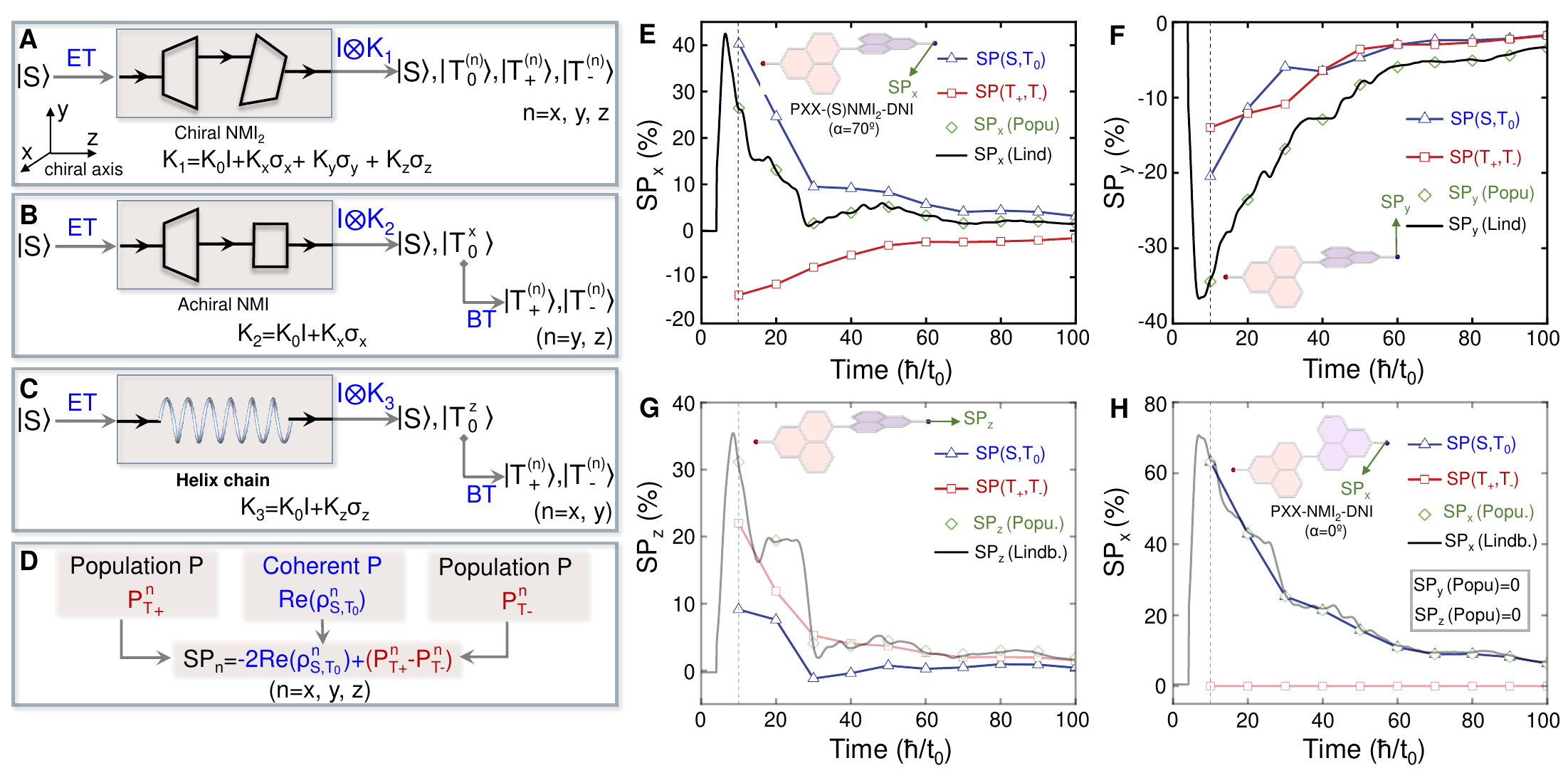}
\caption{\textbf{Triplet-state evolution and the equivalence between population polarization and vector spin polarization in donor–chiral bridge–acceptor molecules.} (\textbf{A-C)}) Comparison of singlet and triplet population evolution for electron transfer through an axially chiral bridge, an achiral bridge, and a one-dimensional helical-chain structure. (\textbf{D}) Schematic illustration of the composition and calculation of the total population polarization. (\textbf{E-G}) Time evolution of the singlet–triplet coherence polarization, triplet population polarization, total population polarization, and the corresponding vector-spin polarization along the three orthogonal directions for the chiral PXX–(S)NMI$_2$–DNI molecule. (\textbf{H}) Time evolution of the same four polarization components along the $x$-direction for the achiral molecule ($\alpha$=0).}
\label{fig4}
\end{figure*}

\noindent{\bfseries Unified physical picture of vector polarization and population polarization}\\
\noindent Within the spin-dynamics framework of spin dynamics, the TREPR signal in the photoinduced charge-separated state D$^{\bullet+}$-B$\chi$-A$^{\bullet-}$ reflects population polarization, the asymmetric occupancy of Zeeman-split levels ($N_{\uparrow}-N_{\downarrow}$). This differs fundamentally from the vector polarizations $\rm SP_y$, $\rm SP_y$, and $\rm SP_z$ computed earlier, which are expectation values of spin angular momentum and report on spatial spin orientation rather than level-population asymmetry. To connect these two descriptions, we adopt a density-matrix formulation. Experimental analyses of chiral systems often invoke a two-electron singlet–triplet basis. Starting from the photoinduced singlet, we map the single-electron transport from donor to acceptor through a completely positive and trace-preserving (CPTP) channel K, whose action is determined by the Lindblad equation incorporating geometric SOC and electron–electron magnetic coupling. Six-state quantum tomography reconstructs the channel’s effect on the Pauli basis \{I, $\sigma_x$, $\sigma_y$, $\sigma_z$\}. Applying this single-electron channel as a tensor product to the donor hole (first spin) while acting only on the acceptor electron (second spin), the initial singlet
\begin{equation}
\rho_S^{(0)} = |S\rangle\langle S| = \frac{1}{4}\left(\rm I\otimes I - \sum_i \sigma_i \otimes \sigma_i\right),
\end{equation}
maps to the unnormalized radical-pair state
\begin{equation}
\tilde{\rho}_{\mathrm{RP}} = (\rm I\otimes {K}) \rho_S^{(0)} = \frac{1}{4}\left[ I\otimes {K}(I) - \sum_n \sigma_n \otimes {K}(\sigma_n) \right],
\end{equation}
which, after normalization, yields $\rho_{\mathrm{RP}}$. From $\rho_{\mathrm{RP}}$, we extract the populations of $\{|S\rangle$, $|T_0\rangle$, $|T_+\rangle$, and $|T_-\rangle\}$. For an axially chiral bridge with $\rm {K} = K_0 I + \sum_n {K}_n \sigma_n$, triplet polulations and $S$-$T_0$ coherences are generated along all three axes, reflecting the three-dimensional nature of population polarization (Figure 4A). In the achiral case, $\rm {K} = K_0 + {K}_x \sigma_x$, producing coherences only along $x$ (Figure 4B); in the one-dimensional helical model, $\rm {K} = K_0 + {K}_z \sigma_z$, so transverse polarizations are suppressed (Figure 4C). The total population polarization along each axis is
\begin{equation}
\rm SP_n^{\mathrm{Popu}} = SP(S,T_0^n) + SP(T_+^n, T_-^n),
\end{equation}
where $\rm SP(S,T_0^n) = -2\,\mathrm{Re}(\rho_{S,T_0}^n)$ and $\rm SP(T_+^n, T_-^n) = P_{T+}^n - P_{T-}^n$ (Figure 4D). This channel-mapping procedure thus embeds both vector and population polarizations within a single density-matrix framework, explicitly linking single-electron transport to the two-electron spin-state observables detected by TREPR.

Figures 4E–4G show the time evolution of the coherence polarization $\rm SP(S,T^n_0)$, the triplet population polarization $\rm SP(T_+^n, T_-^n)$, and the total population polarization $\rm SP_n^{\mathrm{Popu}}$ for the PXX–(S)NMI$_2$–NDI molecule along the perpendicular directions ($n$=$x$, $y$) and chiral axis ($n$=$z$) directions. The total population polarization in each direction is numerically equal to the corresponding vector spin polarization: $\rm SP_n^{\mathrm{Popu}}(t)$=$\rm SP_n(t)$, confirming the consistency between the density-matrix decomposition and the Lindblad master-equation results. Several features emerge from the decomposition. First, both coherence and triplet-population terms contribute to the total polarization along all three axes, reflecting the three-dimensional character of population polarization. Notably, the transverse (perpendicular to the chiral axis) components arise intrinsically from the axial chirality, without requiring transfer from the axial direction via extrinsic mechanisms, consistent with the robust perpendicular TREPR signals observed in D-B$\chi$-A molecules. Second, the three directions exhibit marked anisotropy. Along $x$-direction, $\rm SP(S, T^x_{0})$ is positive and dominates the total, while $\rm SP(T^x_{+}, T^x_{-})$ is negative and smaller. Along $y$-direction, both contributions are negative and comparable, with coherence dominating only at early times. Along $z$-direction, the triplet population term $\rm SP(T^z_{+}, T^z_{-})$ is substantially larger than the coherence term, reversing the hierarchy seen in $x$-direction.

To isolate the chirality dependence of the individual population-polarization components, we repeat the decomposition for the achiral planar configuration ($\alpha$=0) (Figure 4H). In this case, the total population polarization survives only along x and is strictly equal to the vector polarization $\rm SP_x$; the y and z components vanish entirely. The nonzero $x$-direction polarization originates solely from the $S$-$T_0$ coherence term  $\rm SP(S, T^x_{0})$, while the triplet population imbalance $\rm SP(T^x_{+}, T^x_{-})$ is zero. Thus, the coherence polarization reflects the in-plane geometric SOC of the achiral framework and is chirality-independent, whereas the triplet population terms—which vanish in the absence of chirality—are uniquely associated with the CISS effect. This assignment aligns with the two-electron singlet–triplet analysis.

Three conclusions follow from the above analysis. First, the anisotropic polarization generated after photoinduced electron transfer comprises a strictly chirality-dependent axial component and a perpendicular component that contains both chirality-correlated ($\sim$30–40\%) and chirality-independent fractions, a decomposition quantitatively matching the two-source TREPR spectra. Second, the chirality-independent fraction originates from the geometric SOC along the twisted bridge pathway, whose magnitude suffices to account for the observed polarization rates. In contrast, the conventional radical-pair mechanism (RPM), relying on the $g$-factor difference $\Delta g$ ($\sim$10$^{-3}$-10$^{-4}$ in organic radicals) and hyperfine coupling ($\sim$10-50 MHz), cannot produce appreciable polarization on the nanosecond timescale, as both are orders of magnitude smaller than the thermal energy and Zeeman splitting; the negligible isotope effect upon H/D exchange further corroborates this assessment. Third, the geometric SOC framework developed here unifies the two polarization sources within a single theoretical description, providing a coherent physical picture of the CISS effect in axially chiral systems.\\

\noindent{\bfseries Discussion}\\
\noindent We have developed a time-dependent quantum dynamics model that explicitly incorporates the atomic structure of the binaphthyl-type chiral bridge dimer in the metal-free PXX–(S,R)NMI$_2$–NDI system, and have used it to unravel the mechanistic origin of the CISS effect in axially chiral molecules. Four central results emerge. First, the geometric SOC generated along the twisted electron pathway exceeds the intrinsic SOC of light atoms by one to two orders of magnitude, providing a sufficient driving force for spin polarization. Second, spin polarization comprises two distinct sources: a chirality-independent component and a chirality-dependent component that exists both along and perpendicular to the chiral axis; the transverse CISS signal is intrinsic to the axial chirality, not a transfer from the axial direction. Third, the non-Abelian curvature correction defines the orientation of the chiral axis, with the second-order geometric phase from non-collinear SOC rotations between the two NMI units playing a central role. Fourth, using a quantum-channel mapping within a unified density-matrix framework, we establish the strict equivalence between vector spin polarization and population polarization, thereby linking single-electron transport to two-electron spin-state observables in a single theoretical scheme. 

Systematic variation of the dihedral, in-plane twist, and folding angles of the chiral bridge further yields a complete geometric phase diagram for the CISS effect, in which the polarization direction of the non-Abelian term provides a mathematical definition of the chiral axis. The calculated polarization components, relative proportions, and chirality dependence agree quantitatively with the TREPR measurements. Notably, the observed spin-polarization efficiency can be accounted for without invoking the radical-pair mechanism. The geometric-SOC framework thus offers a coherent physical picture that resolves longstanding ambiguities in the interpretation of CISS in chiral molecules and provides explicit structure–property guidelines for the design of chiral materials in spintronic and quantum-information applications.\\

\noindent{\bfseries ASSOCIATED CONTENT}

\noindent{\bfseries Supporting Information}\\
The Supporting Information is available free of charge at
https://pubs.acs.org/doi/***/***.

\begin{adjustwidth}{2em}{0pt} 
 Geometric SOC, theoretical method to calculate spin polarizations and others.
\end{adjustwidth}
\vspace{0.5\baselineskip}
\noindent{\bfseries AUTHOR INFORMATION}

\noindent\hspace*{-0.3em}\ACSSISubtitle{Corresponding Author}
\begin{adjustwidth}{1.5em}{0pt} 
\textbf{Hua-Hua Fu} -- 
\textit{School of Physics and Wuhan National High Magnetic Field Center,
Huazhong University of Science and Technology, Wuhan 430074, People's Republic of China;
Institute for Quantum Science and Engineering, Huazhong University of Science and Technology, Wuhan 430074, People's Republic of China;} %
orcid.org/0000-0003-3920-6324;
Email: \href{mailto:hhfu@hust.edu.cn}{hhfu@hust.edu.cn}.
\end{adjustwidth}
\vspace{0.5\baselineskip}

\noindent\hspace*{-0.3em}\ACSSISubtitle{Authors}
\begin{adjustwidth}{1.5em}{0pt}

\textbf{Shu-Zheng Zhou} --
\textit{School of Physics and Wuhan National High Magnetic Field Center,
Huazhong University of Science and Technology, Wuhan 430074, People's Republic of China.} %

\noindent
\textbf{Xi Sun} --
\textit{School of Physics and Wuhan National High Magnetic Field Center,
Huazhong University of Science and Technology, Wuhan 430074, People's Republic of China.} %

\noindent
\textbf{Kai-Yuan Zhang} --
\textit{School of Physics and Wuhan National High Magnetic Field Center,
Huazhong University of Science and Technology, Wuhan 430074, People's Republic of China.} %

\end{adjustwidth}

\vspace{0.5\baselineskip}

\noindent\hspace*{-0.5em}\ACSSISubtitle{Notes}\\
The authors declare no competing financial interest.\\

\noindent{\bfseries ACKNOWLEDGEMENTS}

\noindent This work is supported by authors' personal resource.\\





\noindent{\bfseries MODELS AND METHODS}\\

\noindent{\bfseries Tight-binding Hamiltonian model}\\
\noindent The total tight‑binding Hamiltonian of the full system includes the donor/acceptor on‑site energies, the on‑site energies and nearest‑neighbor hoppings of the chiral bridge atoms, interface couplings, and the geometric SOC term:
\begin{equation}
\mathcal{H} = \mathcal{H}_D + \mathcal{H}_A + \mathcal{H}_{B\chi} + \mathcal{H}_{D-B\chi} + \mathcal{H}_{B\chi-A} + \mathcal{H}_{\rm geo}.
\end{equation}
Here $\mathcal{H}_{D(A)}$ = $\varepsilon_{D(A)} c_{D(A)}^\dagger c_{D(A)}$, with $\varepsilon_{D(A)}$ the donor (acceptor) on‑site energy. Upon photoexcitation, the donor produces a high‑energy electron and the acceptor receives it, so we set $\varepsilon_{D}$>$\varepsilon_{A}$, with the difference determined by the excitation photon energy. The chiral bridge part is $\mathcal{H}_{B\chi}$ = $\sum_{n=1}^N \sum_{\sigma=\uparrow,\downarrow} \varepsilon_n c_{n\sigma}^\dagger c_{n\sigma}+ \sum_{\langle mn\rangle}\sum_\sigma t_{mn} (c_{m\sigma}^\dagger c_{n\sigma} + {\rm H.c.})$, where $\varepsilon_n$ is the on‑site energy at site $n$, and $t_{mn}$ is taken uniformly as $t_0$. The donor–bridge and bridge–acceptor interface hopping terms are $\mathcal{H}_{D-B\chi}$ = $\tau_D (c_D^\dagger c_{B_1} + {\rm H.c.})$, $\quad \mathcal{H}_{B\chi-A}$ = $\tau_A (c_A^\dagger c_{B_n} + {\rm H.c.})$, with $\tau_D$ = $\tau_A$ = $0.4 t_0$ to reflect weak coupling; $B_1$ and $B_n$ denote the bridge terminal atoms connected to the donor and acceptor, respectively.

The photoinduced charge transfer is a non‑equilibrium open quantum process, which we simulate using a Lindblad‑type master equation:
\begin{align*}
\frac{d\rho}{dt} = &-\frac{i}{\hbar}[\mathcal{H},\rho]+\frac{\Gamma}{\hbar} \mathcal{D}[L_{D\to B1}]+\rho\frac{\Gamma}{\hbar} \mathcal{D}[L_{BN\to A}]\\
&+\rho\frac{\Gamma_{1\to 26}}{\hbar} \mathcal{D}[L_{1\to 26}]+\rho\frac{\Gamma_d}{\hbar} \sum_{n,\mu} \mathcal{D}[L_{n\mu}]\rho,
\end{align*}
where $\rho$ is the density matrix and $\mathcal{D}[L]\rho = L\rho L^\dagger - \frac{1}{2}(L^\dagger L\rho + \rho L^\dagger L)$ is the Lindblad dissipative superoperator. The unidirectional reflectionless transfer operators describe, respectively, electron transfer from donor to the first bridge atom ($L_{D\to B_1} = \sum_\sigma |B_1,\sigma\rangle\langle D,\sigma|$), from the last bridge atom to acceptor ($L_{B_N\to A} = \sum_\sigma |A,\sigma\rangle\langle B_N,\sigma|$), and along a specific intra‑bridge pathway ($L_{1\to 26}$=$\sum_\sigma |26,\sigma\rangle\langle 1,\sigma|$). Local dephasing is described by $(L_{n\mu}$=$|n\rangle\langle n|$$\otimes$$\sigma_\mu)$ ($\mu$=$x,y,z$) with strength $\Gamma_d$. Initially, the electron is localized on the donor in an unpolarized mixed state $\rho(0)$=$\frac{1}{2}(|D,\uparrow\rangle\langle D,\uparrow|$+$|D,\downarrow\rangle\langle D,\downarrow|)$, satisfying total probability 1 and zero initial polarization. Time evolution is performed with a fourth‑order Runge–Kutta method using a time step $\Delta t = 0.1\ {\rm ps}$ and total duration $T_{\max}$=$100\ {\rm ps}$. The on‑site energies are set as follows: $\varepsilon_D$=$5t_0$, for the left half of the bridge (atoms 1–14) $\varepsilon_n$=$2t_0$, for the right half (atoms 15-28) $\varepsilon_n$=0, and $\varepsilon_A$=$-2t_0$. To analyze the migration pathways of electrons among bridge atoms, we define the instantaneous Hamiltonian probability current $J_{i\to j}^H(t) = \frac{2}{\hbar} {\rm Im}\, {\rm Tr}_s [\mathcal{H}_{ij}\rho_{ji}(t)]$, which can be extracted from the density‑matrix dynamics to compute current densities on each chemical bond (detailed derivation in the Supplementary Information). The above framework is directly employed for subsequent numerical calculations of spin‑polarization dynamics and quantitative comparison with experimental results.\\

\end{document}